\documentclass[useAMS]{mn2e}

\usepackage{amsmath}
\usepackage{graphicx}

\title[Improvements on analytic modelling of stellar spots]{Improvements on analytic modelling of stellar spots}
\author[M. Montalto et al.]{M. Montalto$^{1}$\thanks{E-mail: Marco.Montalto@astro.up.pt}, 
G. Bou\'e$^{3}$, M. Oshagh$^{1,2}$, I. Boisse$^{4}$, G. Bruno$^{4}$, N. C. Santos$^{1,2}$\\
$^{1}$Centro de Astrof\'isica da Universidade do Porto, (CAUP), 4150-762, Porto, Portugal\\
$^{2}$Departamento de F\'isica e Astronomia, Faculdade de Ciencias, Universidade do Porto, R. do Campo Alegre, 4169-007, Porto, Portugal\\
$^{3}$Department of Astronomy and Astrophysics, University of Chicago, 5640 South Ellis Avenue, Chicago, IL, 60637, USA\\
$^{4}$Aix Marseille University, CNRS, LAM (Laboratoire d'Astrophysique de Marseille) UMR 7326, 13388, Marseille Cedex 13, France\\
}

\begin{document}

\date{}

\pagerange{\pageref{firstpage}--\pageref{lastpage}} \pubyear{2013}

\maketitle

\label{firstpage}

\begin{abstract}
In this work we present the solution of the stellar spot problem using
the   Kelvin-Stokes theorem.  Our result  is  applicable for any given
location  and  dimension  of the spots  on   the stellar  surface.  We
present explicitely the result up  to the  second  degree in the  limb
darkening law.   This technique     can be  used to    calculate  very
efficiently mutual photometric effects   produced by eclipsing  bodies
occulting stellar spots and to  construct complex spot shapes. 
\end{abstract}


\section{Introduction}

Analytic modelling of photometric  variations induced by stellar spots
was presented several years  ago (Budding 1977) providing  the general
solution for any given spot  dimension and location across the stellar
disk and any  given limb darkening law.  This solution holds under the
assumptions that the star is  a sphere and the  spot profile is drawn
from the interception of a cone with the stellar sphere.

Subsequently other investigators faced this problem from a theoretical
standpoint. Dorren  (1987) and Eker (1994)  for example, reobtained an
equivalent solution to that one of  Budding (1977), but adopted a more
convenient  set of integration    variables.   In other cases     more
restrictive  solutions than  the  one  of  Budding  (1977)  have  been
discussed (e. g. Kipping 2012 ).

One of the most important limitations of current analytic models is 
their inability to construct spot regions being limited to the case
of the so called circular spots introduced by Budding (1977).
The purpose of this work is to reobtain the result of Budding (1977) by
following  a different method, by  means of which we   will be able to
construct the solution for more  complex and realistic spots shapes in
a very  efficient and strightforward  way.   Our results  can be  also
applied to  the calculation of  photometric effects due to  the 
occultation of spot regions by transiting objects.

It is worth noting that  photometric spot modelling is also  performed
by means of numeric integrations on a pixellized grid (e. g.  Lanza et
al. 2010, Boisse et al. 2007, Oshagh et al. 2013).
Nonetheless, despite  the  computational power of modern  computers is
far superior to those available in the past, analytic techniques still
play a very important role in the context of stellar activity studies.
This is due to  the fact that  inverting a lightcurve to  derive spots
geometries and distributions is in fact a very complex problem that in
general does not lead to a unique solution.  It is therefore necessary
to explore  a huge parameter space to  derive  the family of solutions
that more likely satisfy the observational constraints.

The dependencies among the physical  parameters are explicitely stated
in analytic  models  given that the   functional form of the  model is
explicitely  given, and  this renders the   analysis of the degenerate
scenarios well suited for this technique.
Moreover analytic techniques allow us to  sample the parameter space at
a considerable  higher  speed than numerical   methods  which could be
subsequently applied only to local  minima regions eventually relaxing
the simplifing assumptions to which analytic methods may be bounded.

It should  be stressed that, in particular  in the case of single-band
purely rotationally modulated  lightcurves (e.  g.   in the absence of
additional information    coming from   eclipses,   line   broadening,
multi-color photometry) it is in general impossible to reconstruct the
true spot distribution on the stellar surface (Russell 1906).  This is
due to  the fact that  both limb  darkening and foreshortening effects
suppress high order terms in the Fourier  expansion of the lightcurve.
By applying the   maximum entropy reconstruction   method it has  been
deduced  that rotationally  modulated    lightcurves generated by   an
arbitrarily complex spot  distribution can be  reproduced by  using no
more than two compact regions (e. g.  Collier Cameron 1997).

Eclipse mapping of stellar spots can be  regarded as a more favourable
approach to reconstruct spot geometries, in  particular in the case of
transiting  planets given that  planets are  small, intrinsically dark
and proximity effects can be neglected.

Photometric  spot  modelling is  also  complementary  to spectroscopic
techniques like  Zeeman  Doppler  Imaging (ZDI)  and  Doppler imaging.
Both of them are applied to the case of  fast rotators.  ZDI allows to
study stellar magnetic fields topologies from  the analysis of sets of
rotationally modulated circularly  polarized  profiles of photospheric
lines   (e.  g.  Donati \&   Brown   1997).  Doppler imaging  monitors
spectral line asymmetries to  reconstruct the spatial  distribution of
active regions (Vogt \& Penrod 1983).

Improvements  on  analytic modelling is highly  reccomended especially
once the observations require higher precision calculations. Thanks to
the advent of  space satellites like MOST (Walker  et al. 2003), CoRoT
(Auvergne et al.  2009), and more recently Kepler (Koch et al. 2010) a
large database of very precise stellar  photometric time-series is now
available.  Perhaps  one  of the limitations  of the  work  of Budding
(1977) is that the solution is explicitely stated  only up to the case
of linear limb  darkening.  For higher  degrees of  limb darkening the
solution  may be obtained from  the   lower degree  solution by  using
recursive relations.  In this  work we decided to  present explicitely
the solution up to the quadratic term in the limb darkening law, since
this is presently the most widely  adopted limb darkening law. Another
limitation  of the analysis of  Budding (1977) is that the calculation
is based on a direct integration of the stellar flux over the $entire$
visible projected  surface of the spot.   If we wish,  for example, to
construct more  realistic spot shapes by  adding several smaller spots
together,  the  solution   presented by  Budding  (1977)  requires  to
subtract the flux from  the overlapped regions  and requires a complex
calculation of the interception geometry.  The same problem is present
to  calculate  the  photometric effect  produced,  for example,  by  a
transiting planet occulting a stellar spot.

As it  has been  recently demonstrated (P\'al  2012)  for the  case of
mutual  transiting  planets, this  problem can  be  greatly simplified
applying   the  Kelvin-Stokes theorem  to  the  occulting  region, and
thereby performing the integral only over  the border rather than over
the entire surface of the region. In this  work we develop further the
idea proposed by P\'al (2012) applying it to the case of stellar spots
and  afterwords  provide   the link    between  the planet   and  spot
calculations.  As   we  will demonstrate  this   approach  is far more
efficient than the     direct integration once   dealing  with complex
geometries  and  still preserves the accuracy   of the solution in its
entire generality.

\section[]{Definition of the problem}
\label{s:definition}

\begin{figure*}
\includegraphics[width=8.5cm]{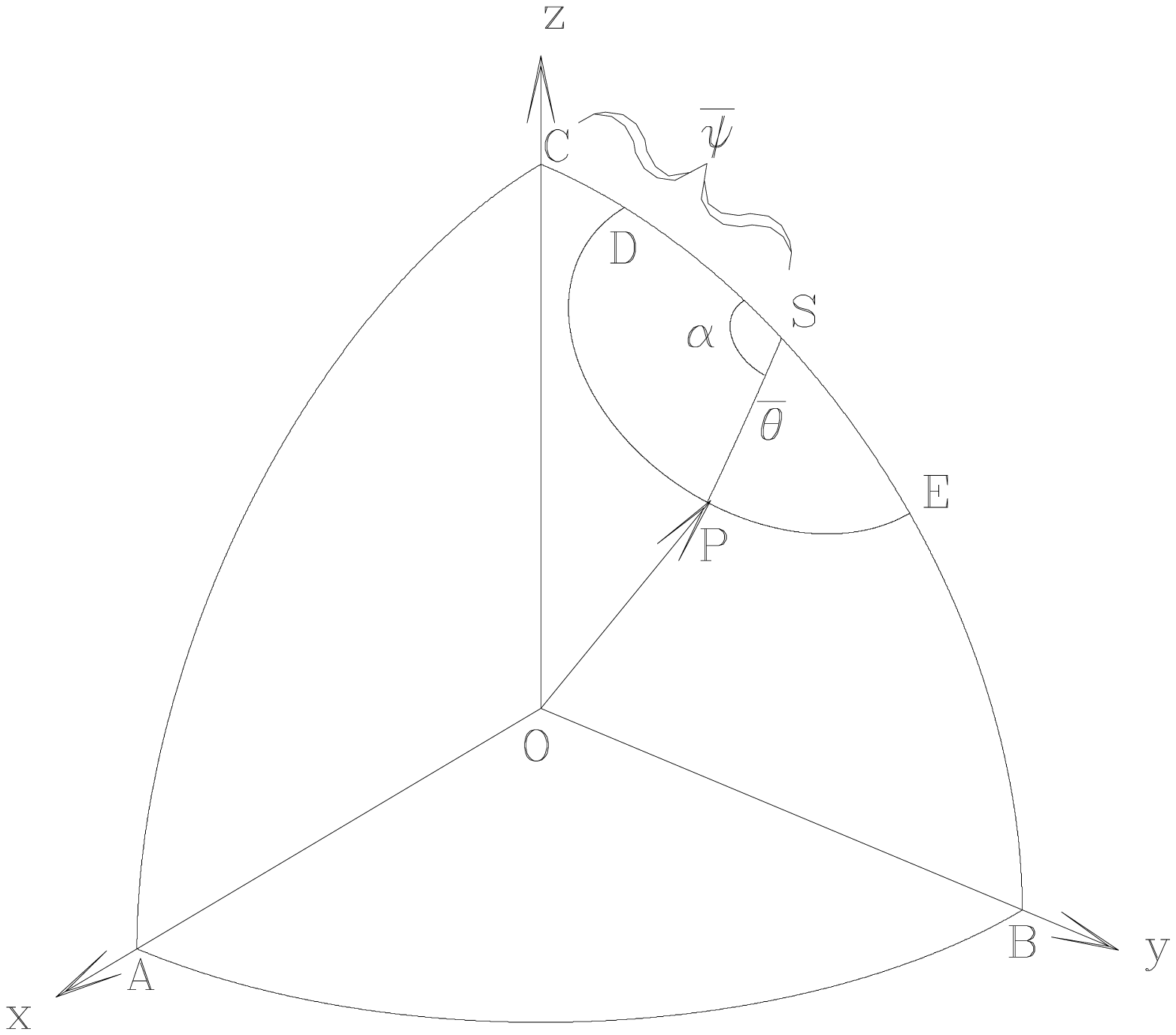}
\includegraphics[width=8.5cm]{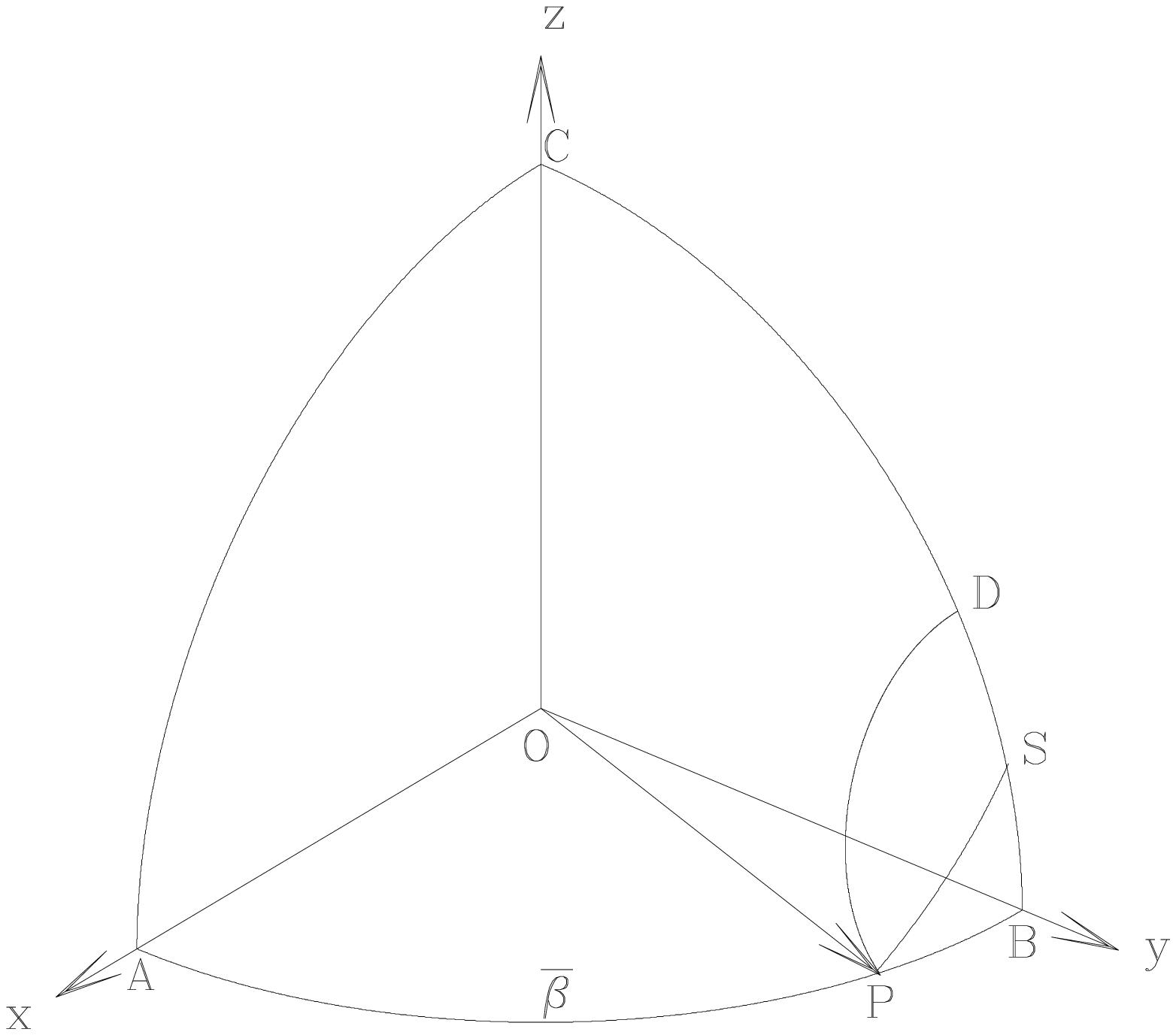}
\caption{ 
Left:  geometry of the problem.  The  observer is  along the $z$ axis,
while  the  $yz$ plane is chosen  correspondent  to the meridian plane
splitting the spot in two identical hemispheres  one of which is shown
in the figure.  The arcs \t{DP},
\t{SP} and  \t{CS} correspond to  $\alpha$, $\overline{\theta}$  and
 $\overline{\psi}$ throughout this work.   Right: geometry at the spot
 intersection with the plane of the sky. The arc \t{AP} in this figure
 corresponds to $\overline{\beta}$ throughout this work.
}
\label{f:fig1}
\end{figure*}

Considering Fig.~1  (left   panel), we define   a cartesian coordinate
system  centered on the stellar sphere  which  radius is normalized to
one.  The $z$ axis of the coordinate system is oriented along the line
of sight  of the observer, and the  $yz$ plane  is coincident with the
meridian of the  star passing through the  center of the stellar spot,
thereby splitting the spot in two  identical hemispheres. Hereafter we
will overline  quantities that  are   considered constant  during  the
integration.   We  define  the   angle $\overline{\psi}$   along  this
meridian, between the sub-stellar point C and the center of the spot S
(the arc \t{CS} in Fig.~1).  Then we consider  a great circle on the
sphere passing through the center of the spot S  and a generic point P
on  the spot profile.   The arc \t{SP}   measured along this circle is
denoted by $\overline{\theta}$.  We  assume here that the spot profile
on the sphere is  drawn from the interception  of a cone (which vertex
is at the center of the sphere) with  the sphere itself, and therefore
$\overline{\theta}$ is constant  denoting the angular dimension of the
spot.   The angle between D and  P in Fig.~1,  measured along the spot
profile, is identified by $\alpha$.

\noindent
Therefore,   by  using  spherical   trigonometry,  we   estabilish the
following   relationships between  the  cartesian  coordinates  of the
position vector  OP, measured  from the center  of  the sphere  to the
generic point P along the spot profile, and the angles above defined

\begin{align*}
x&=\sin\bar{\theta}\sin\alpha\\
y&=\sin{\overline{\psi}}\cos\bar{\theta}-\cos\overline{\psi}\sin\bar{\theta}\cos\alpha\\
z&=\cos\overline{\psi}\cos\bar{\theta}+\sin\overline{\psi}\sin\bar{\theta}\cos\alpha,\\
\end{align*}

\noindent
and for the first derivatives of OP with respect to $\alpha$ we have

\begin{align*}
\frac{dx}{d\alpha}&=\sin\bar{\theta}\cos\alpha\\
\frac{dy}{d\alpha}&=\cos\overline{\psi}\sin\bar{\theta}\sin\alpha\\
\frac{dz}{d\alpha}&=-\sin\overline{\psi}\sin\bar{\theta}\sin\alpha.
\end{align*}

\noindent
In Fig~1 (right panel) we present the geometry of the problem once the
spot intercepts the plane of the sky. In this case we define the angle
$\overline{\beta}$ between the position vector  OP at the interception
point  and the $x$ axis   of the cartesian coordinate system  measured
along the great circle  on the plane of  the sky. Recurring again to
spherical  trigonometry  we estabilish  the following   definition for
$\overline{\beta}$

\begin{align}
\nonumber
\sin\overline{\beta}=\frac{\cos\overline{\theta}}{\sin\overline{\psi}}
\end{align}

\noindent
Thanks to the Kelvin-Stokes theorem  we know that the surface integral
of the curl  of  a vector field   $\bmath{F}$ across a  closed surface
$\bmath{S}$ on the sphere is equal  to the line intgral of $\bmath{F}$
along the border $\bmath{\Sigma}$ of the surface

\begin{equation}
\int_S \nabla \times \bmath{F} \cdot \bmath{dS} = \oint_{\Sigma} \bmath{F} \cdot \bmath{d\Sigma}.
\end{equation}

\noindent
If we assume for $\bmath{F}$ the following functions\footnote{The reader
will note that the functions $\bmath{F}$ we adopted are concident with
those  presented  in   P\'al   (2012) with   the  only   exception  of
$\bmath{F_0}$ for which we assumed a  slightly simpler expression.  It
is however strightforward to  calculate the correspondent integral  of
P\'al (2012) using our choice for $\bmath{F_0}$.}

\begin{align}
\bmath{F_0}=x\bmath{j},
\end{align}

\begin{align}
\bmath{F_1}=-y\frac{1-(1-x^2-y^2)^{\frac{3}{2}}}{3(x^2+y^2)}\bmath{i}+x\frac{1-(1-x^2-y^2)^{\frac{3}{2}}}{3(x^2+y^2)}\bmath{j},
\end{align}

\begin{align}
\bmath{F_2}=\Big(-\frac{1}{2}x^2y-\frac{1}{6}y^3\Big)\bmath{i}+\Big(\frac{1}{6}x^3+\frac{1}{2}xy^2\Big)\bmath{j},
\end{align}

\noindent
where $\bmath{i}$ and $\bmath{j}$ are the unit vector along the $x$ and $y$ axis,
the respective curl evaluated on the surface of the sphere are

\begin{align}
\nabla\times\bmath{F_0}=\bmath{k}
\end{align}

\begin{align}
\nabla\times\bmath{F_1}=z\bmath{k}
\end{align}

\begin{align}
\nabla\times\bmath{F_2}=(x^2+y^2)\bmath{k}.
\end{align}

\noindent
where $\bmath{k}$  is the unit vector along  the $z$ axis.  

Adopting a quadratic  limb  darkening law  and denoting  with $\mu$ as
usual the   cosine of the  angle  between the  normal  to the stellar
surface at a  given point and the line   of sight, we can express  the
stellar intensity as
 
\begin{align}
\nonumber
\frac{I(\mu)}{I(1)}&=1-c_1(1-\mu)-c_2(1-\mu)^2=\\
\nonumber
&=(1-c_1-2c_2)+(c_1+2c_2)z+c_2(x^2+y^2)
\end{align}

\noindent
since $\mu=z$.     Therefore,  by using   the  above  definitions  for
$\bmath{F}$  and extending the integral of $\nabla\,\times\,\bmath{F}$
on  the left hand side of  Eq.~1 to the surface   of the spot, we will
obtain  respectively the projected surface of  the spot along the line
of sight,  the       linear   and the quadratic    limb      darkening
corrections. Equivalently, these integrals can  be calculated from the
right hand side of Eq.~1, integrating $\bmath{F}$  along the border of
the spot.   In the next Section  we will  proceed following the second
approach.

\section{The solution}
\label{s:solution}

\subsection{Projected surface}

For the calculation of the projected surface (which corresponds to the
flux  variation under the assumption  of a  constant luminosity across
the stellar disk), the right-hand side of Eq.~1 reads

\begin{align}
\nonumber
I_\Sigma^0(\overline{\alpha})&=\int_{\Sigma}x\bmath{j}\cdot \bmath{d\Sigma} =\\
\nonumber
&=\int_0^{\overline{\alpha}}\sin\overline{\theta}\sin\alpha\cos\overline{\psi}\sin\overline{\theta}\sin\alpha\,d\alpha.
\end{align}

\noindent
The solution is:

\begin{align}
I_{\Sigma}^0(\overline{\alpha})&=\frac{1}{2}\sin^2\overline{\theta}\cos\overline{\psi}(\overline{\alpha}-\sin\overline{\alpha}\cos\overline{\alpha}).
\end{align}

\noindent
Once the spot is fully visible ($\overline{\psi}<\frac{\pi}{2}-\overline{\theta}$) we set $\overline{\alpha}=2\pi$ which gives

\begin{align}
I_{\Sigma}^0(2\pi)&=\pi\sin^2\overline{\theta}\cos\overline{\psi}.
\end{align}

\noindent
Once        the        spot          is       partially        visible
($\frac{\pi}{2}-\overline{\theta}\leq\overline{\psi}\leq\frac{\pi}{2}+\overline{\theta}$)
     we                      set
$\overline{\alpha}=-\cot\overline{\psi}\cot\overline{\theta_0}$    (which
corresponds to  the value of $\alpha$ at  the  interception point with
the plane of the   sky),   and we have     then to proceed with    the
integration along  the   plane of the sky. To    do that we will   use  as
integration  variable  the  more   convenient  angle $\beta$    defined
in Sect.~\ref{s:definition} which  is  measured along the  great  circle
where  the spot interception arc  resides. We also note that $\bmath{d\Sigma}$
can be expressed simply as $(-y\bmath{i}+x\bmath{j})d\beta$. Therefore
the remaining integral is given by

\begin{align}
\nonumber
I_{\Sigma_b}^0(\overline{\beta})&=\int_{\Sigma}x\bmath{j}\cdot(-y\bmath{i}+x\bmath{j})d\beta=\\
\nonumber
&=\int_{\overline{\beta}}^{\frac{\pi}{2}}x^2 d\beta=\int_{\overline{\beta}}^{\frac{\pi}{2}}\cos^2\beta\,d\beta=\\
&=\frac{1}{2}\Big(\frac{\pi}{2}-\overline{\beta}-\sin\overline{\beta}\cos\overline{\beta}\Big),
\end{align}

\noindent
where $\overline{\beta}$ has been defined in Section~\ref{s:definition}. 
To have the result we have finally to sum Eq.~8 and Eq.~10

\begin{align}
I_{\Sigma}^0=2\Big(I_{\Sigma}^0(\overline{\alpha})+I_{\Sigma_b}^0(\overline{\beta})\Big),
\end{align}

\noindent
where the factor   of two  has  been  introduced to account  for   the
remaining symmetrical hemisphere of the spot.

The  procedure reported above can  be applied in  the same manner also
for the  other  cases   corresponding  to linear  and   quadratic limb
darkening. Here we only report the results of the calculations.

\subsection{Linear limb darkening}

By substituting Eq.~3  in the right hand  side of Eq.~1 and proceeding
with the integration we  obtain  the following  result for  the linear
limb darkening correction

\begin{align}
\nonumber
I_{\Sigma}^1(\overline{\alpha})&=\frac{1}{3}\Big(-\frac{1}{8}\sin^2\overline{\psi}\sin2\overline{\theta}\sin\overline{\theta}\sin2\overline{\alpha}-\\
\nonumber
&-\frac{1}{2}\sin\overline{\theta}\sin\overline{\psi}\sin\overline{\alpha}\Big[\cos(\overline{\psi}-2\overline{\theta})+\\
\nonumber
&+\cos(\overline{\psi}+2\overline{\theta})-2\cos\overline{\theta}\Big]-\\
\nonumber
&-\frac{1}{32}\overline{\alpha}\Big[3\cos(2\overline{\psi}-3\overline{\theta})-8\cos(\overline{\psi}-2\overline{\theta})-\\
\nonumber
&-3\cos(2\overline{\psi}-\overline{\theta})-3\cos(2\overline{\psi}+\overline{\theta})-\\
\nonumber
&-8\cos(\overline{\psi}+2\overline{\theta})+3\cos(2\overline{\psi}+3\overline{\theta})+\\
\nonumber
&+16\cos\overline{\psi}+30\cos\overline{\theta}+2\cos(3\overline{\theta})\Big]+\\
\nonumber
&+2\arctan\Big[\tan\frac{\overline{\alpha}}{2}\cos\Big(\frac{\overline{\theta}+\overline{\psi}}{2}\Big)\sec\Big(\frac{\overline{\psi}-\overline{\theta}}{2}\Big)\Big]+\\
&+\cos\overline{\psi}\sin^2\overline{\theta}\,\overline{\alpha}-\sin\overline{\psi}\cos\overline{\theta}\sin\overline{\theta}\sin\overline{\alpha}\Big).
\end{align}

\subsection{Quadratic limb darkening}

For the quadratic limb darkening, with the aid of the following  
useful definitions

\begin{align}
\nonumber
A_1&=-(\frac{1}{2}+\cos^2\overline{\psi})(\sin^3\overline{\theta}\cos\overline{\theta}\sin\overline{\psi})\\
\nonumber
A_2&=(\frac{1}{2}\sin^4\overline{\theta}\cos\overline{\psi})(1+\cos^2\overline{\psi})\\
\nonumber
A_3&=\frac{1}{2}\sin^2\overline{\theta}\cos^2\overline{\theta}\cos\overline{\psi}\sin^2\overline{\psi}\\
\nonumber
A_4&=-\frac{1}{6}\cos^3\overline{\theta}\sin\overline{\theta}\sin^3\overline{\psi}\\
\nonumber
A_5&=\frac{1}{2}\cos^2\overline{\theta}\sin^2\overline{\theta}\sin^2\overline{\psi}\cos\overline{\psi}\\
\nonumber
A_6&=-\frac{1}{2}\cos\overline{\theta}\sin^3\overline{\theta}\sin\overline{\psi}\cos^2\overline{\psi}\\
\nonumber
A_7&=\frac{1}{6}\sin^4\overline{\theta}\cos^3\overline{\psi}\\
A_8&=\frac{1}{6}\sin^4\overline{\theta}\cos\overline{\psi}
\end{align}


\noindent
and

\begin{align}
\nonumber
B_1&=\frac{\sin^3\overline{\alpha}}{3}\\
\nonumber
B_2&=\frac{1}{32}(4\overline{\alpha}-\sin4\overline{\alpha})\\
\nonumber
B_3&=\frac{1}{2}(\overline{\alpha}-\sin\overline{\alpha}\cos\overline{\alpha})\\
\nonumber
B_4&=\sin\overline{\alpha}\\
\nonumber
B_5&=\frac{1}{2}(\overline{\alpha}+\sin\overline{\alpha}\cos\overline{\alpha})\\
\nonumber
B_6&=\frac{1}{12}(9\sin\overline{\alpha}+\sin3\overline{\alpha})\\
\nonumber
B_7&=\frac{1}{32}(12\overline{\alpha}+8\sin2\overline{\alpha}+\sin4\overline{\alpha})\\
B_8&=\frac{1}{32}(12\overline{\alpha}-8\sin2\overline{\alpha}+\sin4\overline{\alpha})
\end{align}

\noindent
we can express the solution as

\begin{align}
I^{2}_{\Sigma}(\overline{\alpha})=\sum_{i=1}^{8}A_i\,B_i
\end{align}

\noindent
As it is    possible to see,  the  solutions  for  the  integrals  are
expressed as simple algebric and inverse trigonometric functions. This
is a general result that was already proven by Budding (1977).

\subsection{Recurrence relations}

As stated in P\'al (2012) it is interesting to calculate the solutions
of the  integrals  also for  any polynomial expansion  of  the stellar
intensity with respect  to  the $x$  and $y$ coordinates.  This can be
done by using recursive relations giving the solution for higher order
terms as a function of lower order terms. Also, as correctly specified
by P\'al (2012)  to obtain the solution it  is sufficient to calculate
the      expressions  for the     integrals      of  the  form   $\int
x^p\,y^q\cos\alpha$ and $\int x^p\,y^q\sin\alpha$.
 
\noindent
First of all  introducing a notation  similar to P\'al  (2012) we note
that in our case we can write

\begin{equation}
M^{'}_{pq}:=x^q\,y^p=(\sin\overline{\theta}\,\sin\alpha)^{q}(\sin\overline{\psi}\,\cos\overline{\theta}-\cos\overline{\psi}\,\sin\overline{\theta}\,\cos\alpha)^{p}
\end{equation}

\noindent
If we adopt the following definitions

\begin{equation}
y_0=0
\end{equation}
\begin{equation}
x_0=\sin\overline{\psi}\cos\overline{\theta}
\end{equation}
\begin{equation}
r=a\,\sin\overline{\theta}
\end{equation}
\begin{equation}
a=-\cos\overline{\psi}
\end{equation}

\noindent
then we can express $\rm M^{'}_{pq}$ as

\begin{equation}
M^{'}_{pq}:=\frac{1}{a^{q}}(x_0+r\cos\alpha)^{p}(y_0+r\sin\alpha)^{q}=\frac{1}{a^q}M_{pq}
\end{equation}

\noindent
where   $\rm  M_{pq}$   is     the   same  expression  as   in   P\'al
(2012). Therefore  Eq.10-Eq.17 of P\'al  (2012)  can be applied simply
substituting   $\rm   M_{pq}$ with  $\rm  \frac{1}{a^{q}}\,M_{pq}$ and
considering the  above definitions.   The  only  case that  cannot  be
treated in this way is once $\overline{\psi}=\frac{\pi}{2}$ (once only half of the
spot surface is visible). In that case however $M^{'}_{pq}$ reduces to

\begin{equation}
M^{'}_{pq}:=\cos^{p}\overline{\theta}\sin^{q}\overline{\theta}\sin^{q}\alpha.
\end{equation}

\noindent
Then for the integrals $\int M^{'}_{pq}\cos\alpha$ and $\int M^{'}_{pq}\sin\alpha$
we have

\begin{equation}
(q+1)\int M^{'}_{pq}\cos\alpha=(\sin\alpha)^{q+1}\cos^{p}\overline{\theta}\sin^{q}\overline{\theta}\\
\end{equation}

\begin{equation}
(q+1)\int M^{'}_{pq}\sin\alpha=-\sin^{q}\alpha\cos\alpha+q\sin\overline{\theta}\int M^{'}_{pq-1}\\
\end{equation}

\section{Integrals on any generic reference system rotated around
the line of sight}

The integrals provided in Sect.\ref{s:solution}  are calculated in the
reference system of each spot having the $y$  axis passing through the
center of  the spot, as  defined in  Fig.1. It  is possible however to
provide some more general expressions to  calculate the integrals in a
generic reference  system, arbitrarily rotated  around the z axis with
respect to the reference system of a spot.


\noindent
Assuming therefore that the $X$ positive axis of this fixed reference
system $XY$ is rotated of an angle $\phi$ (positive counter-clockwise)
with respect to the positive $x$-axis of a spot as defined in Fig.1,
the novel $X$ and $Y$ coordinates of a generic point on the spot
border, as expressed in the new system are

\begin{align}
\nonumber
X&=x\cos\phi+y\sin\phi\\
Y&=-x\sin\phi+y\cos\phi
\end{align}

\noindent
where $x$ and $y$  are the coordinates of  the point in  the reference
system of  the  spot,  presented in Sect.\ref{s:definition}.  The
expressions for the integrals are reported below

\subsection{Projected surface}

For the projected surface, applying the above equations for $X$ and $Y$
and the function $\bmath{F_{0}}$ defined in Eq.~2 we have

\begin{align}
\nonumber
I_{\Sigma}^0(\overline{\alpha},\phi)&=\frac{1}{2}\sin^2\overline{\theta}\cos\overline{\psi}(\overline{\alpha}-\frac{1}{2}\sin\overline({2\alpha})\cos(2\phi))+\\
\nonumber
&+\frac{1}{8}\sin^2\overline{\theta}\sin(2\phi)\cos(2\overline{\alpha})(1+\cos^2\overline{\psi})-\\
\nonumber
&-\frac{1}{2}\sin\overline{\psi}\sin(2\overline{\theta})\sin^2\phi\sin\overline{\alpha}-\\
&-\frac{1}{8}\sin(2\overline{\psi})\sin(2\overline{\theta})\sin(2\phi)\cos\overline{\alpha}
\end{align}

\subsection{Linear limb-darkening}

For the linear limb darkening we simply have



\begin{align}
\nonumber
I_{\Sigma}^1(\overline{\alpha},\phi)&=I_{\Sigma}^1(\overline{\alpha})
\end{align}


\noindent
where $I_{\Sigma}^1(\overline{\alpha})$ has been obtained in Sect.\ref{s:solution}.

\subsection{Quadratic limb-darkening}

Introducing the  following definitions  where $A_i$  and $B_i$  are as
reported in Sect.\ref{s:solution}

\begin{align}
\nonumber
q&=\frac{\sin2\phi\,\cos2\phi}{2}\\
\nonumber
m&=-\cos^2\phi\,\sin^2\phi+\frac{1}{2}(\cos^4\phi+\sin^4\phi)\\
\nonumber
n&=\cos^2\phi\,\sin^2\phi+\frac{1}{6}(\cos^4\phi+\sin^4\phi)\\
\nonumber
A^{'}_1&=\,2\,m\,A_1\\
\nonumber
A^{'}_2&=\,2\,m\,A_2\\
\nonumber
A^{'}_3&=\,2\,m\,A_3\\
\nonumber
A^{'}_4&=\,6\,n\,A_4\\
\nonumber
A^{'}_5&=\,6\,n\,A_5\\
\nonumber
A^{'}_6&=\,6\,n\,A_6\\
\nonumber
A^{'}_7&=\,6\,n\,A_7\\
\nonumber
A^{'}_8&=\,6\,n\,A_8\\
\nonumber
A^{'}_9&=\frac{q}{3}\sin^4\overline{\theta}(1+3\cos^2\overline{\psi})\\
\nonumber
A^{'}_{10}&=\frac{q}{3}\cos^3\overline{\theta}\sin\overline{\theta}\cos\overline{\psi}\sin^3\overline{\psi}\\
\nonumber
A^{'}_{11}&=-q\cos^2\overline{\theta}\sin^2\overline{\theta}\sin^2\overline{\psi}(\cos^2\overline{\psi}+1)\\
\nonumber
A^{'}_{12}&=q\cos\overline{\psi}\sin\overline{\psi}\cos\overline{\theta}\sin^3\overline{\theta}(\cos^2\overline{\psi}+2)\\
\nonumber
A^{'}_{13}&=\frac{q}{3}\cos^2\overline{\psi}\sin^4\overline{\theta}(\cos^2\overline{\psi}+3)\\
\nonumber
A^{'}_{14}&=-q\cos\overline{\psi}\sin\overline{\psi}\cos\overline{\theta}\sin^3\overline{\theta}
\end{align}


\noindent
and

\begin{align}
\nonumber
B^{'}_{i}&=B_i,\,i=1,\,8\\
\nonumber
B^{'}_{9}&=\frac{\sin^4\overline{\alpha}}{4}\\
\nonumber
B^{'}_{10}&=-\cos\overline{\alpha}\\
\nonumber
B^{'}_{11}&=\frac{\sin^2\overline{\alpha}}{2}\\
\nonumber
B^{'}_{12}&=-\frac{\cos^3\overline{\alpha}}{3}\\
\nonumber
B^{'}_{13}&=-\frac{\cos^4\overline{\alpha}}{4}\\
\nonumber
B^{'}_{14}&=\frac{1}{12}(\cos(3\overline{\alpha})-9\cos\overline{\alpha})
\end{align}

\noindent
the solution is expressed as

\begin{align}
I^{2}_{\Sigma}(\overline{\alpha},\phi)=\sum_{i=1}^{14}A^{'}_i\,B^{'}_i
\end{align}

\noindent
where the reader can  verify  that imposing $\phi=0$  all
the above  integrals
reduce to the simpler expressions presented in Sect.\ref{s:solution}.

\noindent
As for the recurrent integrals we note that in the novel reference
system the expression for $M^{'}_{pq}$ is

\begin{align}
\nonumber
M^{'}_{pq}=X^{p}Y^{q}=(x\cos\phi+y\sin\phi)^{p}(-x\sin\phi+y\cos\phi)^{q}.
\end{align}

\noindent
Since we can always expand the above expression as a polynomial of $x$
and  $y$, and since we  already described in Sect.\ref{s:solution} how
to obtain the  solution for  any  polynomial expansion of the  stellar
intensity  as  a function of   the spot   coordinates, to  obtain  the
solution in the  fixed coordinate system it is  sufficient  to sum the
results of  the  integration over   each term   of  the expansion   of
$M^{'}_{pq}$.

\begin{figure*}
\includegraphics[width=7.5cm]{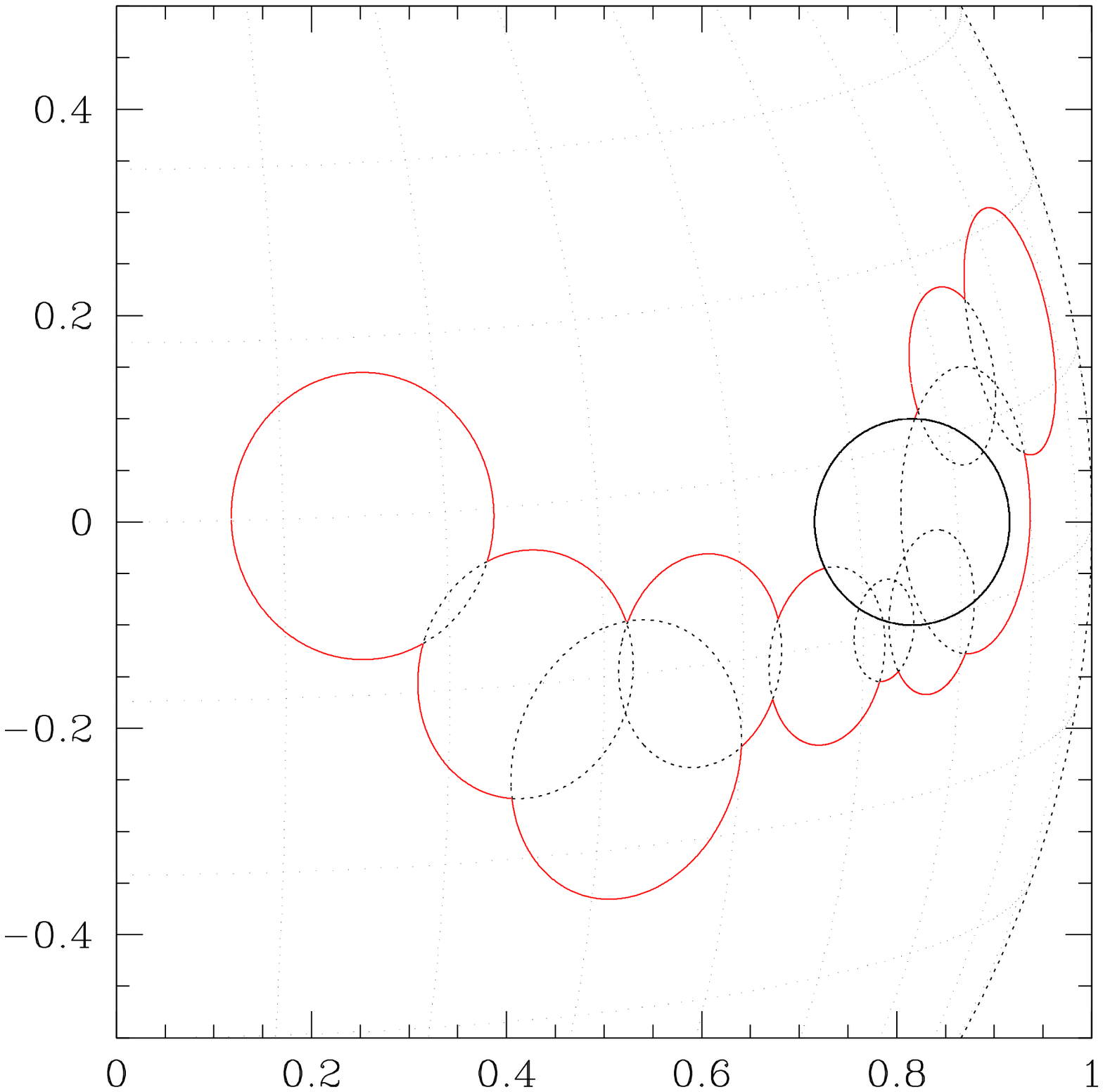}
\includegraphics[width=7.5cm]{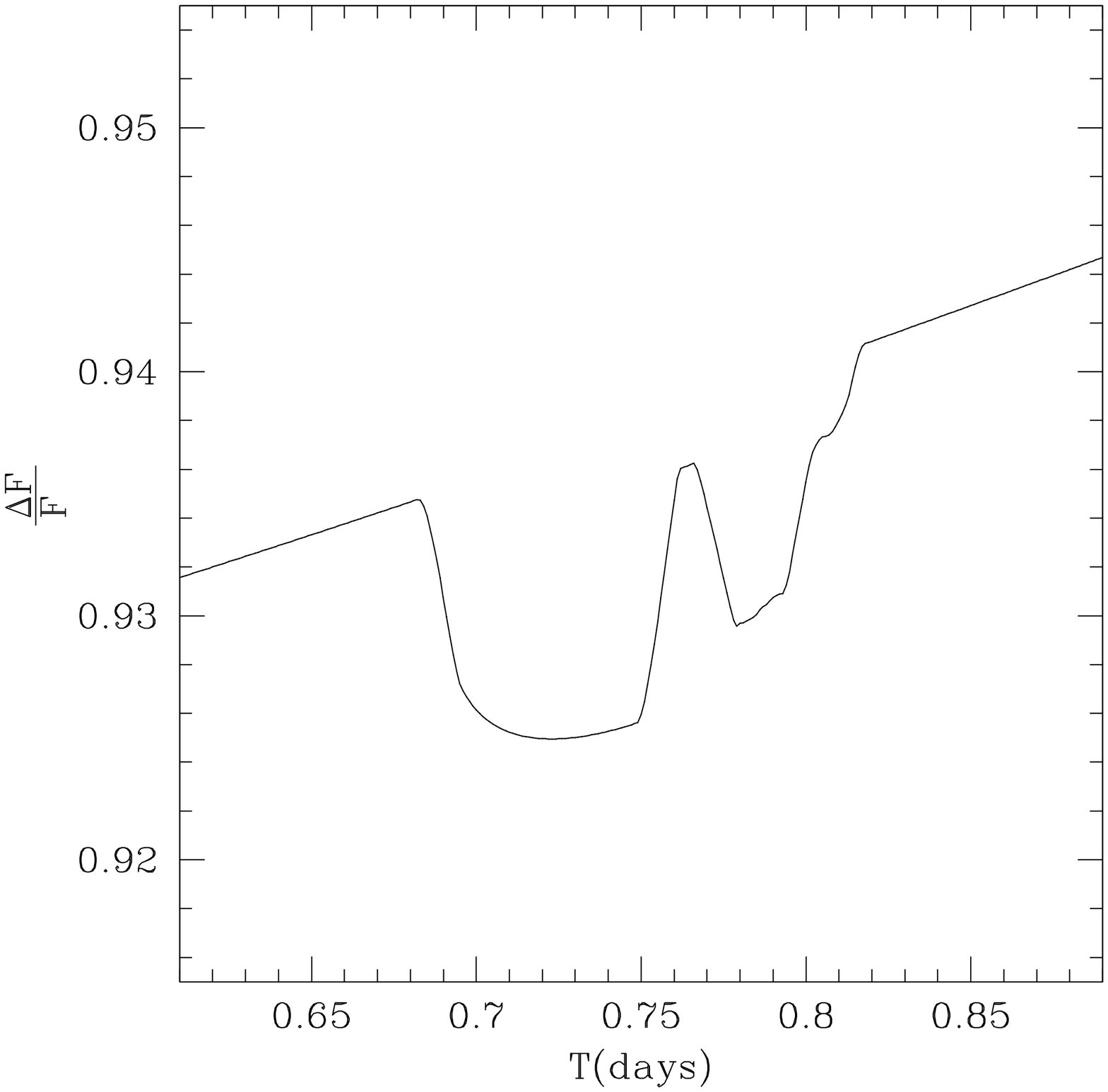}
\caption{ 
Left: the  profile of a  Jupiter planet overlapped  to  a complex spot
region composed by ten spots.  Black denotes integration on the planet
border, red on the spots border.  Dotted lines on the objects profiles
denote invisible  arcs, wherease those on  the stellar  surface denote
stellar meridians and parallels.  X  and Y axis indicate the positions
of the  objects on   the  stellar surface   normalized to  the stellar
radius.  Right: the correspondent  transit lightcurve during which the
overlap of the planets with the  spot region on  the left side occurs,
assuming an  orbital period of 3  days  for the planet  and a rotation
period of 9 days for the star.
}
\label{f:example}
\end{figure*}

\section{Calculation of complex spot structures and occultations of spots by transiting bodies}
\label{s:calculus}

It is now possible  to  provide the  solution for the  calculation  of
multiple spots  overlap
and  occultations   of  spots  by  transiting
bodies. For the latter case we note that P\'al (2012) already provided
the  integrals to calculate the flux  intercepted by a transiting body
with the Kelvin-Stokes theorem.  Given  that in this work we  provided
the  integrals  for the case  of spots  with the same  approach, it is
sufficient to join the two results to obtain the general solution.
 
For convenience we will adopt here the same naming convention of P\'al
(2012) with only a slight useful modification.  The projected shape of
a spherical transiting body on the plane of the sky is a circle, while
that one of a spot is an ellipse as it is  possible to see for example
from our  definitions of $x$  and $y$.  The  interception of  a circle
with  an ellipse yields  in  general a quartic   equation which can be
solved  for each couple of  objects with classic   methods to find the
roots (see below).  The interception points define a set of arcs which
are characterized with the  notation $k:C\{k',l\}$ where the number  k
indicates the circle  or ellipse to which   this l-th arc  belongs and
$C\{k',l\}$ is a  list of circles  or ellipses that  contain this arc.
To  identify if the arc  is an arc of  ellipse or  circle we introduce
here  the  modification that  the  indexes  $k$, $k'$  can   be either
negative  or  positive.  All  the arcs that   define  the  boundary of
integration $\partial S$  are those that  are either corresponding  to
the stellar  boundary  (which we denote with   the index  1),  and are
therefore of the form 1:$\{\}$ or those that  are contained only in it
(k:$\{1\}$). In  the  case in which  we  impose that  the spots have a
different contrast ratio\footnote{ For  simplicity we assume  that all
the overlapping spots  have the same  contrast ratio which corresponds
to the same contrast ratio of the region they  are forming.  }  $f$ to
the stellar surface than  the  planets, (which are considered  totally
dark) we need  to  multiply the integrals  on the  spots arcs  by  the
factor $f$ and we need also to consider in the integration the arcs of
the planets contained inside the spots (see e.g.  Fig.\ref{f:example})
integrating on these  arcs two times: the  first on the  same sense of
the other  planetary arcs that are  not occulting the  spots (to close
the surface of the planet),  and the second  in the opposite sense (to
close the surface of the spot region) multiplying this second integral
by the factor $f$.  Keeping this in mind if  we consider like in P\'al
(2012) a set of  arcs $a$ satisfying the  above conditions which union
is the boundary of integration $a\subset\partial S$ we can express the
total integral in a compact form as

\begin{align}
I=\sum_{a\in\partial S} (I(\phi_a^2)-I(\phi_a^1)),
\end{align}

\noindent
where $\phi_a^1$  and $\phi_a^2$ are the  extreme of integration along
the   generic  arc  a, as   defined  by  the  intersection points, and
$I(\phi_a^2)$, $I(\phi_a^1)$   are   the values   of  the  appropriate
integrals at  these extreme  points.  If  the  arc a  is  denoted by a
negative  index k we   will adopt  our  set of  integrals reported  in
Eq.~26,27,28  (or Eq.~8,12,15).   If the index k  is positive or is
the stellar boudary we  will adopt the respective integrals calculated
by P\'al~(2012).

\noindent
To determine the values of $\phi_a^1$, $\phi_a^2$ located on an arc of
a  spot we proceed  as  follows.   Considering a cartesian  coordinate
system $XYZ$ where the  $X$ axis is now  oriented toward the North and
the $Y$ axis toward the East and the $Z$ axis is always along the line
of sight, the coordinates  of the center of a  spot on the  surface of
the star  will be given by a  longitude angle $\beta_0$  measured from
the $X$ axis toward the positive $Y$ axis, and  by a co-latitude angle
$\overline{\psi}$ measured from the sub-stellar point toward the plane
of the sky. If $X_p$, $Y_p$, $Z_p$ denote the cartesian coordinates of
the interception point  in  this system, the coordintes  $x_p$, $y_p$,
$z_p$ in the reference system of the spot shown in Fig.~1 are obtained
inverting Eq.25,  and by  means of   these coordinates it  is  easy to
obtain  the   correspondent  value  of $\overline{\alpha}$  using  the
definitions in Section~\ref{s:definition}.

%
%

In such a way we can
step from one arc to the other along  the whole boundary to obtain the
solution.

\begin{figure*}
\includegraphics[width=5.0cm]{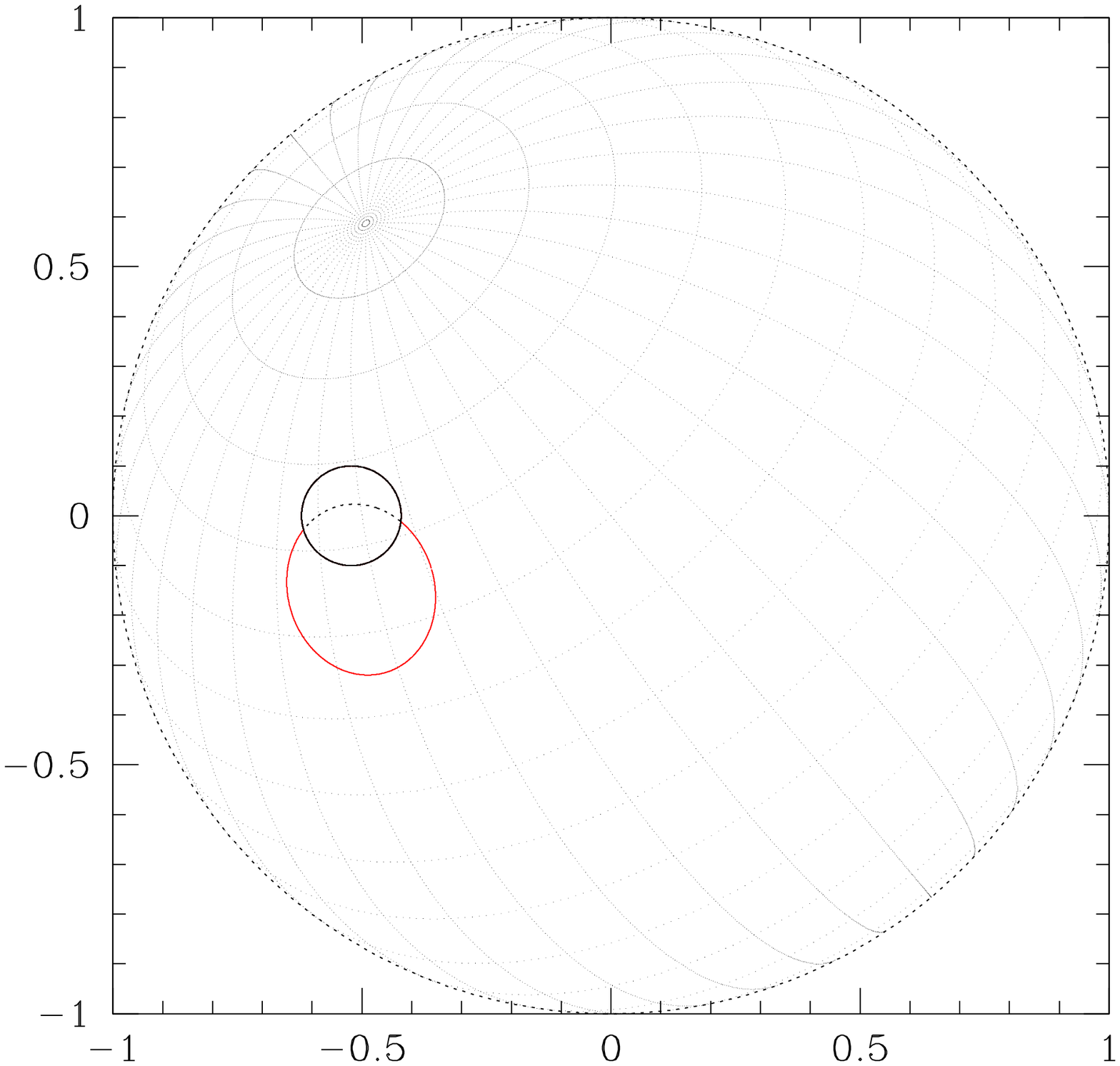}
\includegraphics[width=5.0cm]{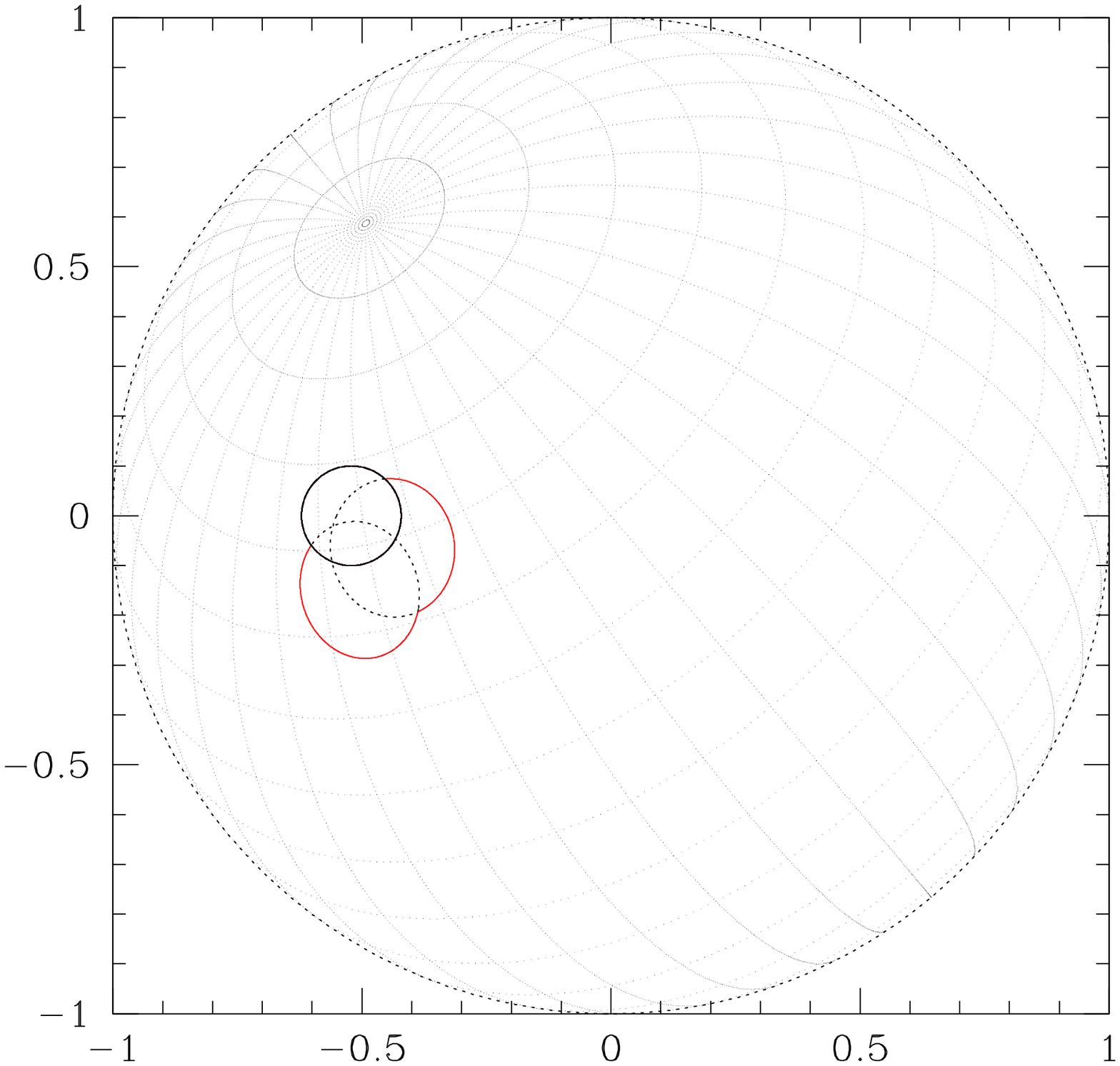}
\includegraphics[width=5.0cm]{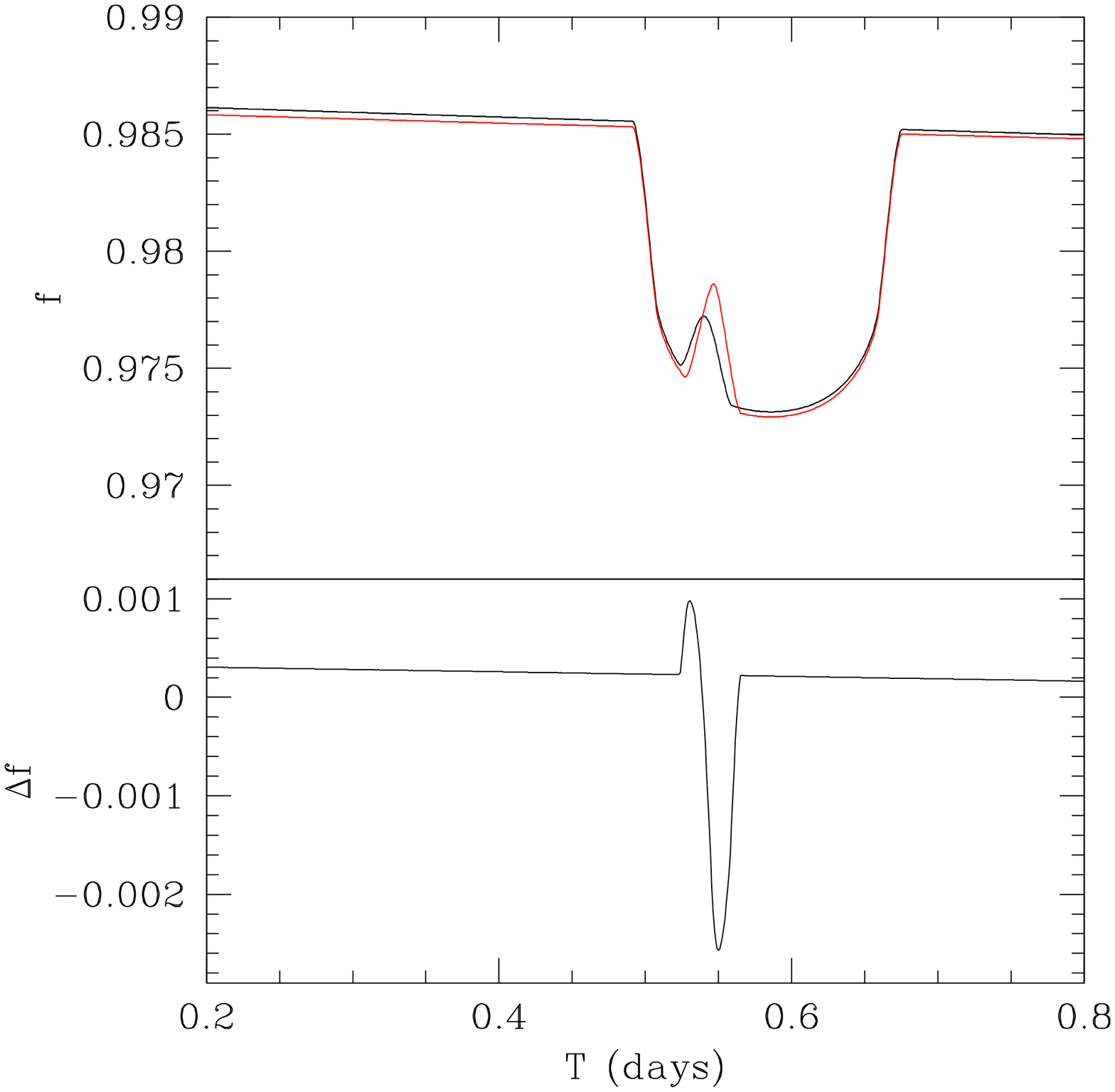}
\caption{ 
On the left side is visible a planet (in black) transiting in front of
a single 10 degrees spot while on the middle panel transiting in front
of  a  spot region composed by  two  8 degrees  spots  offering to the
observer  the same projected  area  of the  single spot. The rightmost
panel shows the two resulting  lightcurves (top) where the single spot
model is denoted by the black  line and the two spot  model by the red
line and the bottom panel shows the difference of the two lightcurves.
}
\label{fig:lc_comparison}
\end{figure*}

\section{Examples}

In  Fig.\ref{fig:lc_comparison},  we present one  application where we
first consider the  overlap of a  single spot,  with angular dimension
equal to 10  degrees  and contrast ratio equal  to  0.5 with a  planet
which radius is equal to 0.1 stellar radii. We additionally considered
a stellar inclination angle  of 50 degrees to the  line of sight and a
position angle (with respect to the y+ axis clockwise) of 320 degrees.
The planet has an orbital period  of 3 days transiting perfectly along
the x-axis in  the same sense of stellar  rotation (P=30  days).  This
configuration produces  the transit   denoted  by the  black line   in
Fig.\ref{fig:lc_comparison}   (rightmost    panel).   Alternatively we
replace in Fig.\ref{fig:lc_comparison} (middle  panel) the single spot
with  a  double spot model   composed by two spots   each of 8 degrees
angular dimension and contrast ratio equal to 0.5, creating overall an
homogeneous spot region.  The disposition of the two spots relative to
each other is selected  in such  a  way that  the projected area  they
offer at the time   of the transit  is  identical to  that one of  the
single  larger spot previously considered.   The  resulting transit is
indicated by  the red  line in  Fig.\ref{fig:lc_comparison} (rightmost
panel), where  in the bottom  panel we  present the difference between
the  two lightcurves. The  result indicates  that  spot  structure may
appreciably   affect the exact  shape  of  the observed  photometric
signal, since  variation of a  few millimagnitudes with respect to the
single spot model  can be expected.  A time  resolution  of around  1 min
would be   sufficient to resolve these  structures.   Depending on the
extension of these  regions spot crossing events may  last also  for a
significant  fraction of the  transit event.  Sub-millimag photometric
precisions can be  acheved  today both from  space and  from the gound
over these timescale.  Studying accurately  these events we can produce
a tomographic analysis of the stellar surface.

\begin{figure}
\includegraphics[width=8.0cm]{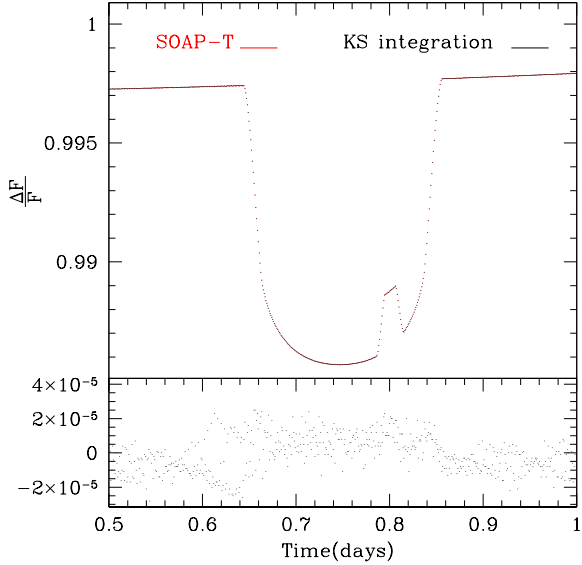}
\caption{ 
Top: A comparison between  $KS$ integration (Kelvin-Stokes integration) and
SOAP-T (Oshagh et  al.~2013) for the case  of  a single circular  spot
crossed by a transiting planet. Bottom: the difference between the SOAP-T
and the $KS$ lightcurve. 
}
\label{fig:comparison_soap}
\end{figure}

As another  example,  we compare in Fig.\ref{fig:comparison_soap}  the
output of $KS$  integration (Kelvin-Stokes integration, black  points)
with the SOAP-T numeric    integrator  result (Oshagh et  al.   2013),
considering the  case of a single circular  spot and a planet crossing
in front of it.   The average difference  over the  considered transit
window   is 8$\times\rm\,10^{-7}$  with a  scatter  equal to $10^{-5}$
denoting a good agreement between the two approaches.

\section{Naked-eye spots}

\begin{figure}
\centering
\includegraphics[width=8.0cm]{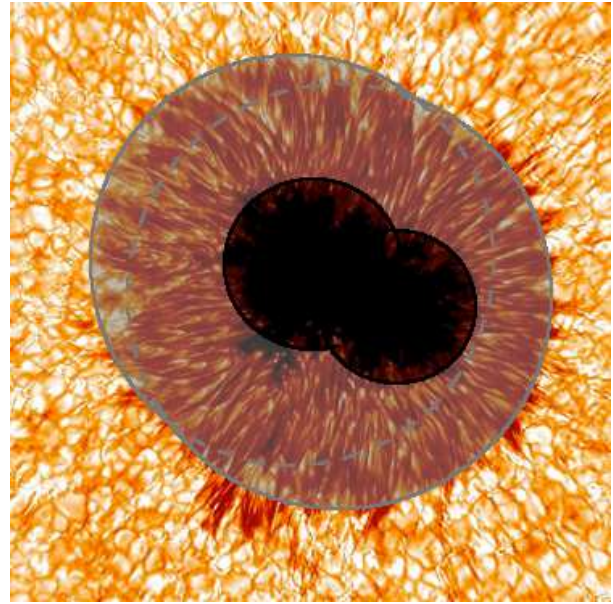}
\caption{ 
In the background is visible an image of a spot region composed by an inner umbra and a surrounding
penumbra (image credit:  Sacramento Peak Observatory of the
National  Solar Observatory  in  New Mexico).  On top of the image  a
tentative reconstruction of  the region using  the method presented  in
this work.
}
\label{fig:spot_image}
\end{figure}

In Fig.\ref{fig:spot_image} we present an image of a spot
region  obtained by the   Sacramento Peak Observatory  of the National
Solar  Observatory in  New Mexico.   On top of the image
we present a possible reconstruction of this region by means of
the  method presented  in  this work. Here   the inner umbra region is
modelled with two overlapping spots of equal contrast ratio, while the
outer penumbra  region is model by  other two  (lower constrast ratio)
overlapping spots. To achieve  the result in  this case is sufficient
to define  some slightly different summation rules  for the spots than
considered  before. For example  spots of equal  contrast ratio can be
summed up homogeneously (thereby   creating a larger spot  region with
the same  contrast  ratio of the   components) whereas for  spots with
different  contrast ratios  we   sum  up  algebrically the    contrast
ratio.  In this  way it  is technically possible  to create  also very
complex regions with  sofisticated  inner-side umbrae and  outer  side
penumbrae.

\section{The quartic}

As explained above the interception  between spot and planet  profiles
leads to a quartic equation for the roots. Here we report the explicit
expression for the quartic  coefficients as a  function of the cosine
of  the  angle $\psi$ between  an interception   point on  the stellar
surface  and   the   line   of   sight.   Defining   $\overline{\psi}$
($\overline{\psi_p}$)  as the  angle between the   center  of the spot
(planet) and the line of sight, and  $\Delta\phi$ as the angle between
the spot and the  planet centers (as seen from  the center of the star
on the plane of the sky), and introducing

\begin{equation}
p=\frac{\sin\overline{\psi}}{\sin\overline{\psi_p}}\sin\Delta\phi
\end{equation}

\begin{equation}
q=\frac{\sin\overline{\psi}}{\sin\overline{\psi_p}}\cos\Delta\phi
\end{equation}

\begin{equation}
\Delta\,r^2=\sin^2\overline{\psi_p}-r^2_p
\end{equation}

\noindent
where $r_p$ is the planet radius we have that  the coefficients of the
quartic are defined as

\begin{align}
A\cos^4\psi+B\cos^3\psi+C\cos^2\psi+D\cos\psi+E=0,
\end{align}

\begin{align}
A=&-\frac{1}{4}(p^2+q^2)\\
B=&q\,\cos(\overline{\psi})\\
C=&\frac{\Delta\,r^2+1}{2}(p^2+q^2)-(\cos^2\overline{\psi}+q\,\cos\overline{\theta})\\
D=&2\cos\overline{\psi}\,\cos\overline{\theta}-q\cos\overline{\psi}(\Delta\,r^2+1)\\
\nonumber
E=&-\frac{p^2\Delta\,r^2}{4}(\Delta\,r^2+2)-\frac{q^2}{4}(\Delta\,r^2+1)^2-\\
 &-\cos^2\overline{\theta}+q\,\cos\overline{\theta}(\Delta\,r^2+1)-\frac{p^2}{4},
\end{align}

\noindent
We recall that four is the highest degree of a polynomial for which an
algebric solution for the roots can  be found.  The solution was first
provided by Lodovico Ferrari (1540).

\section{Conclusions}

In  this work we  presented the  solution of  the stellar spot problem
obtained  applying the Kelvin-Stokes  theorem.  The  solution has been
expressed   in   a  closed form   up   to the   second   degree of the
limb-darkening  law  and recursive  expressions  are provided  for any
polynomial expansion of  the  stellar  intensity.  We  expressed   the
results both in  the reference system  of  the spot  and in  a generic
reference system arbitrarily oriented around the  line of sight of the
observer.  Coupled with previous results on planets presented in P\'al
(2012) we have now a powerful technique  to incorporate the effects of
stellar  activity    into   light-curve  modelling  including   mutual
photometric effects  produced by planets  with planets or planets with
spots and allowing for the creation of complex and more realistic spot
regions, and independent on the   number, positions and dimensions  of
these objects.

The  software implementation of the method  described in this paper is
made publically    available  directly    at the     following address
http://eduscisoft.com/KSINT/    or        through     the         link
http://www.astro.up.pt/exoearths/tools.html. It is written in  Fortran
95  and  it has  been   tested on a  Linux  machine  using  a gfortran
compiler. The program is called  KSint ad can generate the  lightcurve
produced by an arbitrary combination of spots and planets.

\section*{Acknowledgments}
MM  acknowledges the support from FCT  in the  form of grant reference
SFRH/BDP/71230/2010.  This work was supported by the European Research
Council/European  Community  under   the FP7  through   Starting Grant
agreement   number  239953.   NCS  was  supported  by  FCT through the
Investigador FCT contract reference IF/00169/2012 and POPH/FSE (EC) by
FEDER funding through the program "Programa Operacional de Factores de
Competitividade" - COMPETE.

\section*{Bibliography}
Auvergne, M., Bodin, P., Boisnard, L., et al. 2009, A\&A, 506, 411\\
Boisse, I., Bonfils, X., Santos, N. C., A\&A, 545, 109\\
Budding 1977, Ap\&SS, 48, 207\\
Collier Cameron 1997, MNRAS, 287, 556\\
Donati, J. F. \& Brown, S. F. 1997, A\&A, 326, 1135\\
Eker, Z. 1994, ApJ, 420, 373\\
Kipping, D. M., 2012, MNRAS, 427, 2487\\
Koch, D. G., Borucki, W. J., Basri, G., Batalha, N. M., Brown, T. M., Caldwell, D., Christensen-Dalsgaard, J., Cochran, W. D. et al. 2010, ApJL, 713, L79\\
Lanza, A. F., Bonomo, A. S., and Rodon\'o, M 2007, A\&A, 464, 741\\
Oshagh, M., Bou\'e G., Figueira, P., Santos, N. C., Haghighipour, N. 2013, A\&A, 558, 65\\
P\'al 2012, MNRAS, 421, 1825\\
Russell H. N. 1906, ApJ, 24, 1\\ 
Vogt, S. S. \& Penrod, G. D. 1983, PASP, 95, 565\\
Walker, G., Matthews, J., Kuschnig, R., Johnson, R., Rucinski, S., Pazder, J., Burley, G., Walker, A. et al. 2003, PASP, 115, 1023\\

\label{lastpage}

\end{document}